\documentclass[aps,showpacs,preprintnumbers,amsmath,amssymb]{revtex4}
\usepackage{graphics,epsfig}
\usepackage{graphicx}
\usepackage{dcolumn}
\usepackage{bm}

\begin{document}
\baselineskip=0.8 cm
\title{{\bf Wave dynamics of a six-dimensional black hole localized on a tensional
three-brane}}
\author{Songbai Chen}
\email{chsb@fudan.edu.cn} \affiliation{Department of Physics,
Fudan University, Shanghai 200433, P. R. China
 \\ Institute of Physics and  Department of Physics,
Hunan Normal University,  Changsha, Hunan 410081, P. R. China }

\author{Bin Wang}
\email{wangb@fudan.edu.cn} \affiliation{Department of Physics,
Fudan University, Shanghai 200433, P. R. China}

\author{ Rukeng Su}
\email{rksu@fudan.ac.cn}
 \affiliation{China Center of Advanced Science and Technology (World Laboratory),
P.B.Box 8730, Beijing 100080, People¡¯s Republic of China
\\ Department of Physics, Fudan University, Shanghai 200433, P. R. China}

\vspace*{0.2cm}
\begin{abstract}
\baselineskip=0.6 cm
\begin{center}
{\bf Abstract}
\end{center}
We study the quasinormal modes and the late-time tail behavior of
scalar perturbation in the background of a black hole localized on
a tensional three-brane in a world with two large extra
dimensions. We find that finite brane tension modifies the
standard results in the wave dynamics for the case of a black hole
on a brane with completely negligible tension. We argue that the
wave dynamics contains the imprint of the extra dimensions.

\end{abstract}

 \maketitle
\newpage
\vspace*{0.2cm}

The study of wave dynamics outside black holes has been an
intriguing subject for the last few decades (for a review, see
\cite{1,2,3}). We have got the schematic picture regarding the
dynamics of waves outside a spherical collapsing object. A static
observer outside a black hole can indicate three successive stages
of the wave evolution. First the exact shape of the wave front
depends on the initial pulse. This stage is followed by a
quasinormal ringing, which describes the damped oscillations under
perturbations in the surrounding geometry of a black hole with
frequencies and damping times of the oscillations entirely fixed
by the black hole parameters. The quasinormal modes (QNM) is
believed as a unique fingerprint to directly identify the black
hole existence. Detection of these QNM is expected to be realized
through gravitational wave observation in the near future
\cite{1,2}. Despite the potential astrophysical interest, QNM
could also serve as a testing ground of fundamental physics. It is
widely believed that the study of QNM can help us get  deeper
understandings of the AdS/CFT\cite{3,ads}, dS/CFT\cite{ds}
correspondences, loop quantum gravity\cite{loop} and also the
phase transition of black holes \cite{phase} etc. At late times,
quasinormal oscillations are swamped by the relaxation process.
This relaxation is the requirement of the black hole no hair
theorem \cite{4}.

It is of interest to extend the study of wave dynamics on usual
black holes in general relativity to black holes obtained in
string theory, since this could help us to get the signature of
the string. Attempt on this respect has been carried out in
\cite{chen}. String theory predicts quantum corrections to
classical General Relativity, and the Gauss-Bonnet terms is the
first and dominating correction among the others. The
investigation on wave evolutions in the Gauss-Bonnet black holes
has been done in \cite{abdalla}. String theory predicts the
existence of the extra dimension. Recent developments on higher
dimensional gravity resulted in a number of interesting
theoretical ideas such as the brane world concept. The essence of
this string inspired model is that Standard Model fields are
confined to a three dimensional hypersurface, the brane, while
gravity propagates in the full spacetime, the bulk. It has been
argued that the extra dimension could imprint in the wave dynamics
in the branworld black holes\cite{shen,molina}. Besides the wave
dynamics, other arguments for detecting extra dimensions in the
braneworld black hole Hawking radiation have also been proposed
\cite{shen, liu, kanti, park}.

In general it is very hard to obtain exact solutions of
higher-dimensional Einstein's equations describing black holes on
brane with tension. Recently, a metric describing a black hole
located on a three-brane with finite tension, embedded in locally
flat six-dimensional spacetime was constructed in \cite{kaloper}.
In spherically symmetric six-dimensional Schwarzschild gauge, it
is described as
\begin{eqnarray}
ds^2=-(1-\frac{M_{BH}}{4\pi^2M^4_*br^3})dt^2+(1-\frac{M_{BH}}{4\pi^2M^4_*br^3})^{-1}dr^2
+r^2\{d\theta^2+\sin^2{\theta}[d\phi^2+\sin^2{\phi}(d\chi^2+b^2\sin^2{\chi}d\psi^2)]\}
,\label{metric}
\end{eqnarray}
where $b$ measures the deficit angle about the axis parallel with
the three-brane  in the angular direction $\psi$ and is related to
the brane tension $\lambda$ by $b =1-\frac{\lambda}{2\pi M^4_*}$.
$M_{BH}$ is the ADM mass of the usual six-dimensional
Schwarzschild black hole and $ M_*$ is the fundamental mass scale
of six-dimensional gravity. The position of the black hole horizon
is located at
\begin{eqnarray}
 r_h=(\frac{M_{BH}}{4\pi^2M^4_*b})^{\frac{1}{3}}.
\end{eqnarray}
For convenience, we can define $M\equiv
\frac{M_{BH}}{4\pi^2M^4_*}$ and then the Hawking temperature of
the black hole can be written as
$T_{H}=\frac{3}{4\pi}(\frac{b}{M})^{1/3}$. Obviously, the presence
of the brane tension changes the relation between the black hole
mass and horizon radius and leads to the existence of a deficit
angle in the angular direction $\psi$ which means that the
geometry of the metric asymptotes a conical bulk space at large
distance ($r\gg r_H$). When the brane tension $\lambda=0$ (i.e.
$b=1$), this metric reduces to the six-dimensional Schwarzschild
metric.

Finite brane tension modifies the standard results if we compare
with the black hole on a brane with negligible tension. This has
been observed in the semiclassical description of the black hole
decay process\cite{dai}. It has been found that the brane tension
alters significantly the power output of small black holes located
on the brane and the power emitted in the bulk diminishes as the
tension increases. This distinct signature is argued to be
observed in high energy collisions. The question we want to ask is
that whether the wave dynamics of black holes can show signatures
of information about brane tension, whether this could complement
collider searches for these information through Hawking radiation.
This is the main motivation of the present work.

We will concentrate on the scalar perturbation in the background
of the black hole localized on a tensional three-brane in a world
with two large extra dimensions. The equation of motion for the
massless scalar field is described by
\begin{eqnarray}
\frac{1}{\sqrt{-g}}\partial_{\mu}(\sqrt{-g}g^{\mu\nu}\partial_{\nu})
\Psi(t,r,\theta,\phi,\chi,\psi)=0.\label{WE}.
\end{eqnarray}
Seperating the variable
$\Psi(t,r,\theta,\phi,\chi,\psi)=\frac{e^{-i\omega
t}R(r)}{r^2}\Theta(\theta)\Phi(\phi)\Gamma(\chi)\Xi(\psi)$ and
expressing the tortoise coordinate
\begin{eqnarray}
r_*=\int{\frac{1}{1-\frac{M}{br^3}}}dr,
\end{eqnarray}
we can get the wave equation of the scalar field in the metric
(\ref{metric}),
\begin{eqnarray}
\frac{d^2}{dr^2_*}R(r)+[\omega^2-V(r)]R(r)=0,\label{rdir}
\end{eqnarray}
\begin{eqnarray}
&&\frac{1}{\sin^3{\theta}}\frac{d}{d\theta}\bigg(\sin^3{\theta}
\frac{d}{d\theta}\Theta(\theta)\bigg)
+(\eta_4-\frac{\eta_3}{\sin^2{\theta}})\Theta(\theta)=0,\label{angd4}\\
&&\frac{1}{\sin^2{\phi}}\frac{d}{d\phi}\bigg(\sin^2{\phi}\frac{d}{d\phi}\Phi(\phi)\bigg)
+(\eta_3-\frac{\eta_2}{\sin^2{\phi}})\Phi(\phi)=0, \label{angd3} \\
&&\frac{1}{\sin{\chi}}\frac{d}{d\chi}\bigg(\sin{\chi}\frac{d}{d\chi}\Gamma(\chi)\bigg)
+(\eta_2-\frac{\eta_1}{\sin^2{\chi}})\Gamma(\chi)=0,\label{angd2} \\
&&\frac{1}{b^2}\frac{d^2}{d\psi^2}\Xi(\psi)+\eta_1\Xi(\psi)=0,\label{angd0}
\end{eqnarray}
with
\begin{eqnarray}
V(r)=(1-\frac{M}{br^3})\bigg(\frac{\eta_4+2}{r^2}+\frac{4M}{br^5}\bigg)\label{efp}.
\end{eqnarray}
Here parameters $\eta_4$, $\eta_3$, $\eta_2$ and $\eta_1$ denote
the eigenvalues of the equations (\ref{angd4})-(\ref{angd0})
respectively. In terms of quantum theory, they are determined by
four quantum numbers ($L$, $l_2$, $l_1$, $m$) of the system. In
the case $b=1$ (i.e., the brane tension $\lambda=0$), the angular
equations (\ref{angd4})-(\ref{angd0}) reduce to those in the
spherically symmetric cases. Their solutions can be expressed as
the expansion in the Gegenbauer functions  with the eigenvalues
$\eta_4=L(L+3)$, $\eta_3=l_2(l_2+2)$, $\eta_2=l_1(l_1+1)$ and
$\eta_1=m^2$ respectively. In this case, the eigenvalue $\eta_4$
is independent of the angular number $m$ and is defined only by
the quantum number $L$. However when $b\neq 1$, the spherical
symmetry is broken and then $\eta_4$ depends on the quantum
numbers $L$ and $m$.

In order to study the QNM and the late-time tail behaviors of the
external perturbations in the black hole spacetime (\ref{metric}),
the first step for us is to determine eigenvalue $\eta_4$ in the
equation (\ref{efp}). Obviously, the eigenvalue $\eta_1$ in the
equation (\ref{angd0}) is $\frac{m^2}{b^2}$. We restrict to $m>0$
here and rewrite equation (\ref{angd2}) as
\begin{eqnarray}
\frac{1}{\sin{\chi}}\frac{d}{d\chi}\bigg(\sin{\chi}\frac{d}{d\chi}\Gamma(\chi)\bigg)
+(\eta_2-\frac{m^2}{\sin^2{\chi}})\Gamma(\chi)
+\frac{1}{\sin^2{\chi}}(m^2-\frac{m^2}{b^2})\Gamma(\chi)=0.\label{angd1}
\end{eqnarray}
We limit ourselves to the case where the deviation of the
parameter $b$ from unity is very small which is physically
justified for small brane tension. Then the third term on the
left-hand-side of the equation above  can be regarded as a
perturbation. Using the perturbation theory, we have
\begin{eqnarray}
\Gamma(\chi)&=&P^m_l(\cos{\chi})+\gamma S^m_l(\cos{\chi})+O(\gamma^2),\nonumber\\
\eta_2&=&\eta^{(0)}_2+\gamma \eta^{(1)}_2+O(\gamma^2),\label{pe1}
\end{eqnarray}
where $\gamma$ is a dimensionless parameter denoting the
perturbation scale. For convenience, we set $\gamma=1$ throughout
our paper. Substituting the variables (\ref{pe1}) into the angular
equation (\ref{angd1}), we can obtain
\begin{eqnarray}
&&(D_0+\eta^{(0)}_2)P^m_{l_1}(\cos{\chi})=0,\label{zero}\\
&&(D_0+\eta^{(0)}_2)S^m_{l_1}(\cos{\chi})+(D_1+\eta^{(1)}_2)P^m_{l_1}(\cos{\chi})=0,\label{1-order}
\end{eqnarray}
with
\begin{eqnarray}
&&D_0=\frac{1}{\sin{\chi}}\frac{d}{d\chi}\bigg(\sin{\chi}\frac{d}{d\chi}\bigg)
-\frac{m^2}{\sin^2{\chi}},\nonumber\\
&&D_1=\frac{1}{\sin^2{\chi}}(m^2-\frac{m^2}{b^2}).
\end{eqnarray}
From the zeroth order equation (\ref{zero}), we have
\begin{eqnarray}
\eta^{(0)}_2=l_1(l_1+1)\label{ben11}.
\end{eqnarray}
Multiplying equation (\ref{1-order}) by $P^m_{l_1}(\cos{\chi})$
from the left and integrating it over $\chi$, we get
\begin{eqnarray}
\eta^{(1)}_2=\frac{m(2l_1+1)(1-b^2)}{2b^2}\label{ben12}.
\end{eqnarray}
Changing the wave function $P^m_l(x)$ to
$\tilde{C}^{s}_h(x,y)=\frac{d^yC^{s}_{h}(x)}{dx^y}$ (where
$C^{s}_{h}(x)$ is the Gegenbauer function) in the equation
(\ref{pe1}) and repeating the operations above, we can obtain the
eigenvalues of equations (\ref{angd3}) and (\ref{angd4})
\begin{eqnarray}
&&\eta_3=l_2(l_2+2)+\frac{m(2l_2+2)(1-b^2)}{2b^2}, \label{ben2} \\
&&\eta_4=L(L+3)+\frac{m(2L+3)(1-b^2)}{2b^2}. \label{ben3}
\end{eqnarray}
The formula (\ref{ben3}) tells us that when $b\neq 1$ the
eigenvalue $\eta_4$ of the angular equation (\ref{angd4}) depends
not only on the multiple moment $L$, but also on the parameter $b$
and angular number $m$. Moreover, the eigenvalue $\eta_4$
increases with the increase of the parameters $m$ and the brane
tension $\lambda$. The perturbation result is in agreement with
the numerical result obtained in \cite{dai}.

To observe the fundamental QNM of the scalar perturbation in the
background of a six-dimensional black hole on the brane with
tension, we can adopt the third-order WKB approximation.  The
formula for the complex quasinormal frequencies $\omega$ in this
approximation is
\begin{eqnarray}
\omega^2=[V_0+(-2V^{''}_0)^{1/2}\Lambda]-i(n+\frac{1}{2})(-2V^{''}_0)^{1/2}(1+\Omega),
\end{eqnarray}
where
\begin{eqnarray}
\Lambda&=&\frac{1}{(-2V^{''}_0)^{1/2}}\left\{\frac{1}{8}\left(\frac{V^{(4)}_0}{V^{''}_0}\right)
\left(\frac{1}{4}+\alpha^2\right)-\frac{1}{288}\left(\frac{V^{'''}_0}{V^{''}_0}\right)^2
(7+60\alpha^2)\right\},\nonumber\\
\Omega&=&\frac{1}{(-2V^{''}_0)}\bigg\{\frac{5}{6912}
\left(\frac{V^{'''}_0}{V^{''}_0}\right)^4
(77+188\alpha^2)\nonumber\\&-&
\frac{1}{384}\left(\frac{V^{'''^2}_0V^{(4)}_0}{V^{''^3}_0}\right)
(51+100\alpha^2)
+\frac{1}{2304}\left(\frac{V^{(4)}_0}{V^{''}_0}\right)^2(67+68\alpha^2)
\nonumber\\&+&\frac{1}{288}
\left(\frac{V^{'''}_0V^{(5)}_0}{V^{''^2}_0}\right)(19+28\alpha^2)-\frac{1}{288}
\left(\frac{V^{(6)}_0}{V^{''}_0}\right)(5+4\alpha^2)\bigg\},
\end{eqnarray}
and
\begin{eqnarray}
\alpha=n+\frac{1}{2},\;\;\;\;\;
V^{(s)}_0=\frac{d^sV}{dr^s_*}\bigg|_{\;r_*=r_*(r_{p})} \nonumber,
\end{eqnarray}
$n$ is the overtone number.

Setting $M=2$ and substituting the effective potential (\ref{efp})
into the formula above, we can obtain the quasinormal frequencies
of scalar field in the black hole localized on a tensional
three-brane.
\begin{table}[h]
\begin{center}
\begin{tabular}[b]{cccc}
 \hline \hline
 \;\;\;\; $b$ \;\;\;\; & \;\;\;\; $\omega\ \ \ (m=0)$\;\;\;\;  & \;\;\;\;  $\omega \ \ \ (m=1)$\;\;\;\;
 & \;\;\;\; $\omega \ \ \ (m=2)$ \;\;\;\; \\ \hline
\\
0.90& \;\;\;\;\;2.847856-0.380539i\;\;\;\;\; & \;\;\;\;
2.898630-0.380473i\;\;\;\;\; & \;\;\;\;
\;2.948528-0.380412i\;\;\;\;\;
 \\
0.92&2.868797-0.383337i&2.908446-0.383285i&2.947561-0.383237i
 \\
0.94&2.889436-0.386095i&2.918480-0.386057i&2.947236-0.386020i
 \\
0.96&2.909785-0.388814i&2.928706-0.388789i&2.947505-0.388765i
\\
0.98&2.929853-0.391495i&2.939103-0.391483i&2.948323-0.391471i
\\
1.00&2.949650-0.394141i&2.949650-0.394141i&2.949650-0.394141i
\\
 \hline \hline
 $b$  &$\omega \ \ \ (m=3)$ & $\omega \ \ \ (m=4)$ & $\omega \ \ \ (m=5)$\\ \hline
 \\
0.90&2.997595-0.380355i&3.045869-0.380302i&3.093389-0.380252i
\\
0.92&2.986163-0.383190i&3.024271-0.383147i&3.061905-0.383105i
\\
0.94&2.975715-0.385985i&3.003922-0.385952i&3.031867-0.385919i
\\
0.96&2.966185-0.388741i&2.984748-0.388718i&3.003195-0.388696i
\\
0.98&2.957515-0.391459i&2.966678-0.391447i&2.975813-0.391436i
\\
1.00&2.949650-0.394141i&2.949650-0.394141i&2.949650-0.394141i
\\ \hline\hline
\end{tabular}
\end{center}
\caption{The fundamental ($n=0$) quasinormal frequencies of scalar
field in the background of a six-dimensional black hole on the
brane with tension, where $L=5$.}
\end{table}
\begin{figure}[ht]
\begin{center}
\includegraphics[width=6cm]{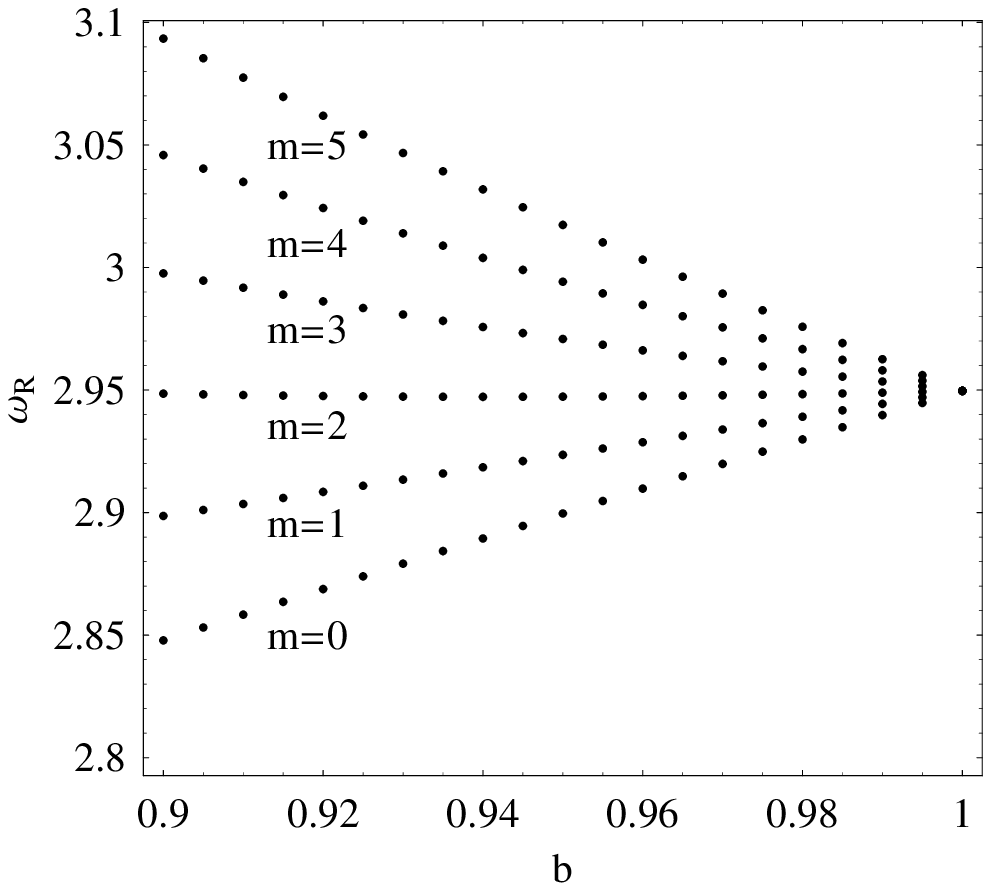}\ \ \ \ \ \ \includegraphics[width=6cm]{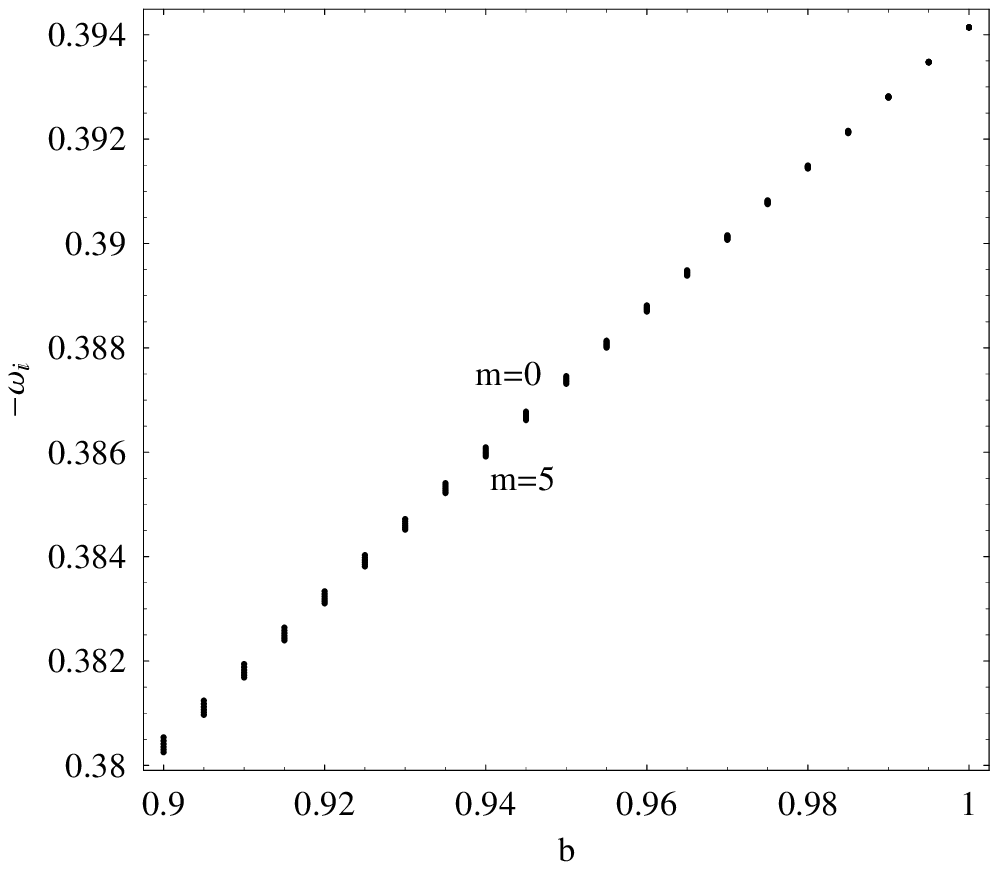}
\caption{Variation of the real part (the left) and the magnitude
of the imaginary part (the right) of quasinormal frequencies with
the parameter $b$ for chosen $L=5$. }
 \end{center}
 \label{fig1}
 \end{figure}
The fundament QNM frequencies for $L=5$ are listed in Table 1. In
Fig.1, we display the variation of the QNM frequencies with brane
tension. For a tensionless brane, where $b=1$, the result
coincides with that of the six-dimensional Schwarzschild black
hole\cite{Cardoso}. Increasing the brane tension, which
corresponds to decreasing the parameter $b$, the absolute value of
the imaginary part of quasinormal frequency decreases.  The
dependence of the real part of quasinormal frequency on the brane
tension is very complicated. For each fixed $L$, the real part of
quasinormal frequency decreases with the increase of the brane
tension for small values of $m$, while increases with the increase
of the brane tension for big values of $m$. For $L=5$, the
property is shown in Fig.1 and the value $m=2$ is the turning
point separating the increasing and decreasing behavior of the
real part frequency with the increase of the brane tension. The
$m$ values' involvement in the dependence of the real part of
quasinormal frequency on the brane tension has been observed for
other chosen $L$. The qualitative property shown in Fig.1 holds
while the turning point value $m$ changes.

We have also showed the dependence of the QNM frequencies on the
angular number $m$ for fixed brane tension. In Fig.2, we see that
the real part frequency increases while the absolute value of the
imaginary part frequency decreases with the increase of $m$ for
fixed brane tension. This has not been observed in the usual six-
dimensional Schwarzschild black hole with $b=1$.

Therefore in this black hole spacetime the quasinormal frequencies
depend not only on the parameter $b$ but also on angular number
$m$. This result can be explained by the fact that due to the
presence of the parameter $b$, the angular eigenvalues $\eta_4$,
the event horizon radius and symmetry of the black hole are
changed from those of the usual spherical cases.

\begin{figure}[ht]
\begin{center}
\includegraphics[width=6cm]{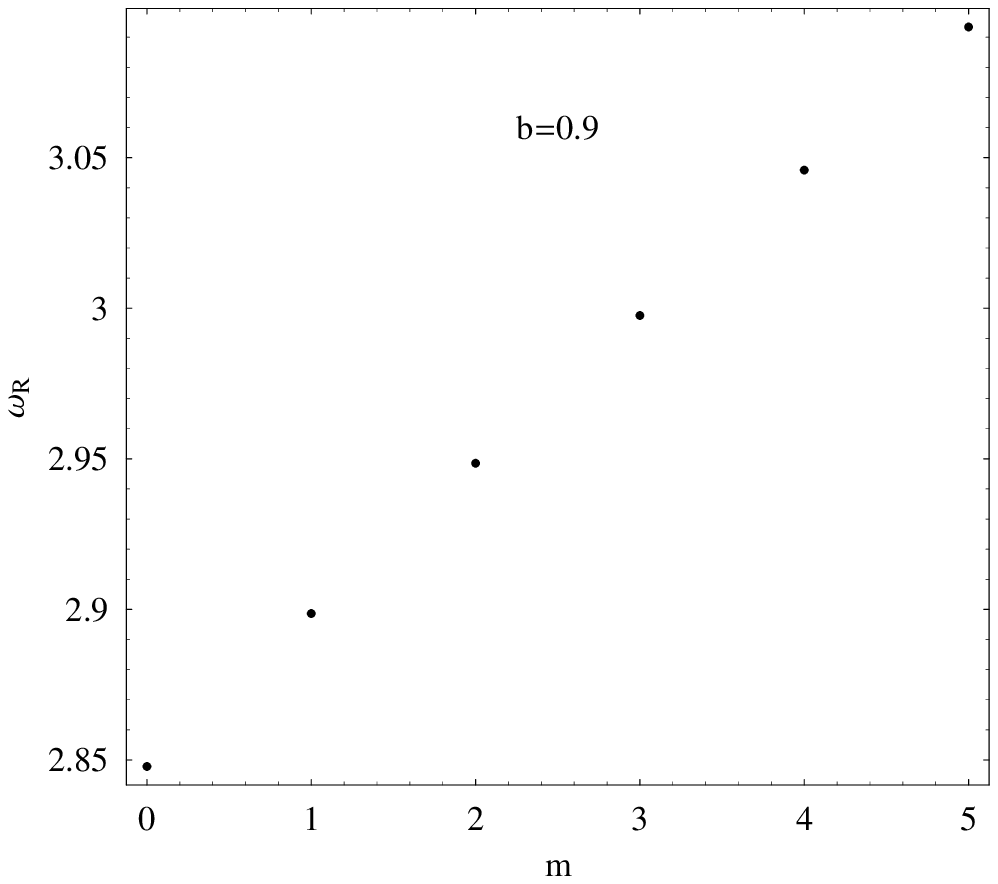}\ \ \ \ \ \
\includegraphics[width=6.3cm]{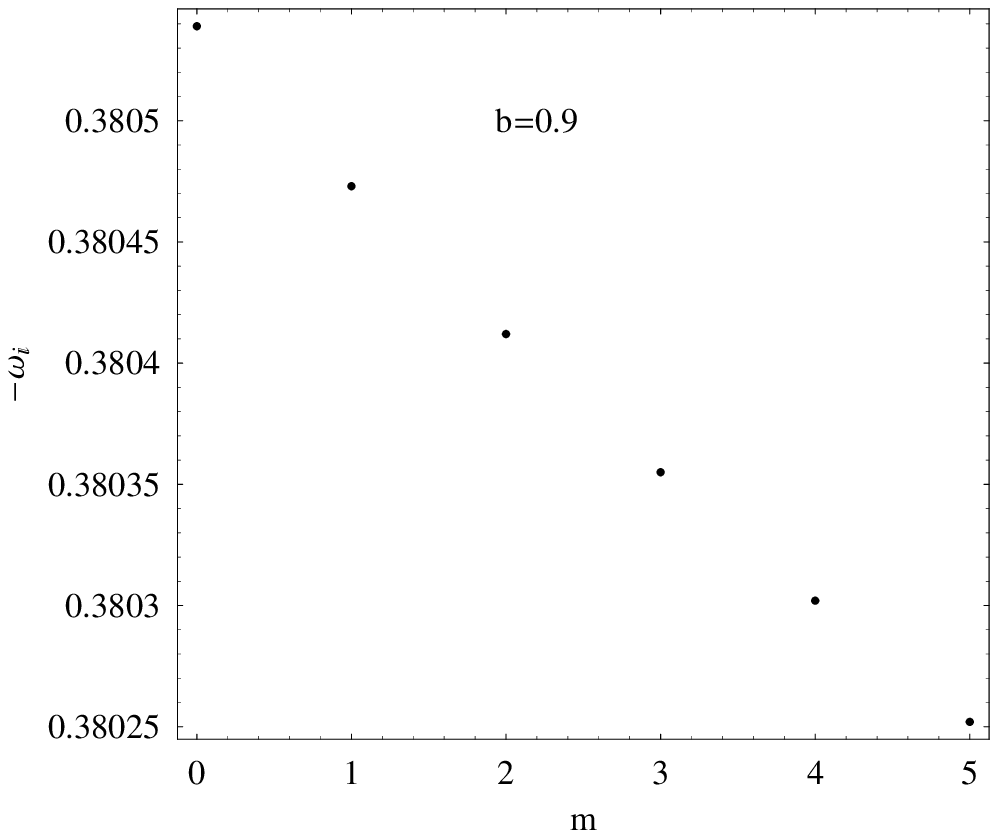}
\caption{Variation of the real part (the left) and the magnitude
of the imaginary part (the right) of quasinormal frequencies with
the angular number $m$ for fixed $L=5$ and $b=0.9$. }
 \end{center}
 \label{fig2}
 \end{figure}

Now we extend our discussion to the late-time tail behavior of the
scalar perturbation in the background of a six-dimensional black
hole on a brane with tension. In Ref.\cite{Ching}, Ching
\textit{et al} made a complete study on the late-time tail in
black hole spacetimes for an external perturbation dominated by
the evolution equation with a potential in the form
\begin{eqnarray}
V(x)\sim \frac{\nu
(\nu+1)}{x^2}+\frac{c_1\log{x}+c_2}{x^{\alpha}},\;\;\;\;\;
x\rightarrow \infty\label{vtail1}.
\end{eqnarray}
After a careful study of the contribution of the branch cut to the
Green's function, they found that the late-time behavior of the
external perturbation in general is dominated  by a power-law or
by a power-law times a logarithm, and its decay rate depends on
the leading term at very large spatial distances. Making use of
their conclusion, Cardoso \cite{Cardoso} studied the late-time
tail of scalar fields in the D-dimensional Schwarzschild spacetime
for the case $c_1=0$. When $\nu$ is an integer, the main
contribution to the late-time tail comes from  the
$\frac{c_2}{x^{\alpha}}$ term and then the form of the tail is
given by
\begin{eqnarray}
&&\Psi\sim t^{-(2\nu+2\alpha-2)}, \;\;\;\;\; \alpha \;\;\text{odd
integer}
<2\nu+3,\label{n1}\\
&&\Psi\sim t^{-(2\nu+\alpha)}, \;\;\;\;\; \;\;\;\;\;\;\text{all
other real}\; \alpha.\label{n2}
\end{eqnarray}
When $\nu$ is not an integer, the late-time tail takes the form
\begin{eqnarray}
\Psi\sim t^{-(2\nu+2)}, \;\;\;\;\; \text{non-integer} \; \nu.
\label{non}
\end{eqnarray}
The reason is that the main contribution to the late-time tail
comes from the $\frac{\nu(\nu+1)}{x^2}$ in this case. In the
background of a six-dimensional black hole on a brane with
tension, the effective potential of the evolution equation for
scalar perturbation can be rewritten as
\begin{eqnarray}
V(r_*)=(1-\frac{M}{br^3})(\frac{\nu(\nu+1)}{r^2}+\frac{4M}{br^5}),
\end{eqnarray}
where
\begin{eqnarray}
\nu=\frac{2L+3}{2}\sqrt{1+\frac{m(1-b^2)}{2(2L+3)b^2}}-\frac{1}{2}.
\end{eqnarray}
When $r_*\rightarrow \infty$, we find that it has the asymptotical
form
\begin{eqnarray}
V(r_*)|_{r_*\rightarrow \infty}=\frac{\nu(\nu+1)}{r^2_*}-\frac{2M
L(L+3)}{br^5_*}\label{vtail}.
\end{eqnarray}
Obviously, this potential is a special case of the potential
(\ref{vtail1}), i.e. $c_1=0$. Thus, comparing with the results in
\cite{Ching,Cardoso}  we can directly obtain the form of the
late-time behavior. For the case when $\nu$ is not an integer, in
terms of equation (\ref{non}) we find that the late-time tail is
described by
\begin{eqnarray}
\Psi\sim
t^{-1-(2L+3)\sqrt{1+\frac{m(1-b^2)}{2(2L+3)b^2}}}.\label{tail1}
\end{eqnarray}
In the case when $\nu$ is an integer, it is easy to obtain that
$2\nu +3>2L+5>5=\alpha$ and the late time tail of the scalar wave
propagation has the form
\begin{eqnarray}
\Psi\sim
t^{-7-(2L+3)\sqrt{1+\frac{m(1-b^2)}{2(2L+3)b^2}}}.\label{tail2}
\end{eqnarray}
From equations (\ref{tail1}) and (\ref{tail2}), we find that in
both cases for the larger multiple moment $L$, angular quantum
number $m$ and brane tension $\lambda$, the decay of the scalar
field at very late time becomes more quickly. The dependence of
the late time tail on the brane tension is consistent with that
observed in the power output result in \cite{dai}. The bigger
brane tension may diminish the power emission of the black hole so
that it is easier for the perturbation outside the black hole to
die out. Moreover, we have also found that the decay rate of
late-time tail in the integer $\nu$ case is larger that in the
non-integer $\nu$ situation.

In summary, we have studied the QNM and the late-time tail
behavior of scalar perturbation in the background of a black hole
localized on a tensional three-brane in a world with two large
extra dimensions. Our results show that with the nonzero brane
tension, the properties of the wave dynamics are different from
those of the usual spherical black holes where the brane tension
is completely negligible. We argue that significant modifications
of the standard wave dynamics due to the presence of the brane
tension compared to that of the black hole on a brane with
negligible tension could serve as signatures, complementing the
information through the evaporation of a black hole off a tense
brane in the future collider searches.

Recently a non-negligible black hole-brane interaction was
discussed in \cite{last}. If the brane tension is small, as taken
in this work, then by exciting degrees of freedom of the scalar
field propagating in the background of a black hole, one would
also excite the degrees of freedom of the brane itself. These
degrees of freedom can also influence the wave dynamics. It would
be interesting to study this influence in detail in the future.

\begin{acknowledgments}

This work was partially supported by NNSF of China, Ministry of
Education of China and Shanghai Education Commission.  S. B.
Chen's work was partially supported  by the National Basic
Research Program of China under Grant No. 2003CB716300.
\end{acknowledgments}

\end{document}